\documentclass[a4paper,10pt]{article}
\usepackage{amssymb,graphicx}
\begin{document}
\vskip 64pt

\begin{center}
{\Large\bf Gravitino Production\\ after Inflation in No-Scale Supergravity} \\
\vspace{16pt}
Kenji Takagi$^{(2)}$, Yuta Koshimizu$^{(1)}$, Toyokazu Fukuoka$^{(1)}$, Takao Sakai$^{(3)}$, \\
Hikoya Kasari$^{(1)}$ and Mitsuo J. Hayashi$^{(1)}$ \\
\vspace{16pt}
$^{(1)}${\it Department of Physics, Tokai University, 1117, Kitakaname, Hiratsuka, 259-1292, Japan} \\
$^{(2)}${\it Company VSN, Minato, Shibaura, 108-0023, Japan} \\
$^{(3)}${\it Company Mobile Techno, Saiwa, Horikawa, 212-0013, Japan}
\vskip 16pt
E-mail: mhayashi@keyaki.cc.u-tokai.ac.jp
\end{center}

\thispagestyle{empty}
\vspace{24pt}

\begin{abstract}
By using a No-Scale Supergravity model, which was proved to explain WMAP observations appropriately, a mechanism of supersymmetry breaking and a preheating just after the end of inflation are investigated. 
Non-thermal production rate of gravitino is estimated as well as supersymmetry breaking mechanism numerically. The supersymmetry breaking is triggered by the inflaton superfield alone and the interchange of supersymmetry breaking fields does not occur in our model. 
By the instant preheating mechanism, the preheating temperature is calculated through the process where the inflaton decays into right handed sneutrinos, which will decay into Higgs fields and others.  The obtained value of the yield variable for gravitino is rather large, however, the primordial gravitinos decay very rapidly and the preheating temperature is lower than the gravitino mass,  the effect to the standard Big Bang Nucleosynthesis (BBN) scenario is negligible. 
Though a tachyonic state seems to appear from modular field $T$, it will be solved by assuming the spontaneous breaking of modular invariance.  Whether or not the fact is a defect, we emphasize that the model still seems phenomenologically effective.
\end{abstract}

\newpage
\section{Introduction}
\addtocounter{page}{-1}

Following ``Five-Year Wilkinson Microwave Anisotropy Probe (WMAP) Observations" \cite{ref:WMAP}, the theory of inflation are proved to be the most promising theory of the early universe before the big bang.

 As a favored scenario to explain the observational data, it is customary to introduce a scalar field called inflaton.
What kind of theoretical frameworks are the most appropriate to describe the theory of particle physics, inflation and the recent observations.
It seems to require a far richer structure of contents than that of the standard theory of particles. 
 As far as the 4D, $N=1$ supergravity can play an elementary role in the theory of  the space-time and the particles \cite{ref:SUGRA}, it can also be essential in the theory of the early universe as an effective field theory. 
Supergravity, however, has been confronting with the difficulties, such as the $\eta$-problem and the supersymmetry breaking mechanism has been studied by many authors \cite{ref:Nilles, ref:NoScale, ref:Witten, ref:Cvetic}.
We have investigated to prevail over these difficulties in Refs.\cite{ref:Hayashi1, ref:Hayashi2} by using the modular invariant supergravity induced from superstring \cite{ref:Ferrara}. We found that the interplay between the dilaton field $S$ and gauge-singlet scalar $Y$ could give rise to sufficient inflation. The model is free from the the $\eta$-problem and realizes appropriate amount of inflation as well as the TT angular power spectrum.  

In supergravity, gravitino is a unique object and cosmological meanings of gravitino is one of the most important problem. In this Letter we will investigate the gravitino production just after the end of inflation. 

First we will briefly review the model and the former results \cite{ref:Hayashi1, ref:Hayashi2} as follows:
It is convenient to introduce the dilaton field $S$, a chiral superfield $Y$ and the modular field $T$. Here, all the matter fields are set to zero for simplicity.
Then, the effective no-scale type K{\"a}hler potential and the effective superpotential that incorporate modular invariance are given by:
\begin{eqnarray*}
&&K=-\ln \left(S+S^\ast\right) -3\ln \left(T+T^\ast-|Y|^2\right), \\
&&W=3bY^3\ln\left[c\>e^{S/3b}\>Y\eta^2(T)\right],
\end{eqnarray*}
where $\eta(T)$ is the Dedekind $\eta$-function, defined by:
\begin{equation}
\eta(T)=e^{-2\pi T/24} \prod^{\infty}_{n=1}(1-e^{-2\pi nT}).
\end{equation}
The parameter $b$ and $c$ are treated as free parameters in this letter as discussed in Ref.\cite{ref:Hayashi2}.
The scalar potential is in order:
\begin{eqnarray}
V(S,T,Y)&=&\frac{3(S+S^\ast)|Y|^4}{(T+T^\ast-|Y|^2)^2}
\Bigg(3b^2 \left|1+3\ln\left[c\>e^{S/3b}\>Y\eta^2(T)\right]\right|^2
\nonumber\\
&&\quad+\frac{|Y|^2}{T+T^\ast-|Y|^2}
\left|S+S^\ast-3b\ln\left[c\>e^{S/3b}\>Y\eta^2(T)\right]\right|^2
\nonumber\\
&&\quad+6b^2|Y|^2\left[2(T+T^\ast)\left|\frac{\eta^\prime(T)}{\eta(T)}\right|^2
+\frac{\eta^\prime(T)}{\eta(T)}
+\left(\frac{\eta^\prime(T)}{\eta(T)}\right)^\ast\right]\Bigg).
\end{eqnarray}
The potential is explicitly modular invariant and can be shown to be stationary at the self-dual points $T=1$ and $T=e^{i\pi /6}$ \cite{ref:Ferrara}.

We had found that the potential $V(S,Y)$ at $T=1$ has a stable minimum at for the values
$b = 9.4$, $c = 131$ and obtained
\begin{eqnarray}
S_{{\rm min}} = 1.51, && Y_{{\rm min}} = 0.00878480,
\end{eqnarray}
where $\eta (1) = 0.768225$, $\eta^2 (1) = 0.590170$, $\eta' (1) = -0.192056$, $\eta'' (1) = -0.00925929$ are used.
The inflationary trajectory can be well approximated by
\begin{equation}
Y_{\rm min}(S) \sim 0.009268 e^{-0.035461 S},
\end{equation}
which corresponds to the trajectory of the stable minimum for both $S$ and $Y$.
The slow-roll parameters $\varepsilon_S$ and $\eta_{SS}$ satisfy the slow-roll conditions.
The the number of $e$-folds $\sim 57$, by fixing the parameters $b=9.4$ and $c=131$ and integrating from $S_{\rm end} \sim 4.19$
to $S_*\sim 11.6$, i.e. our potential can produce
a cosmologically plausible number of $e$-folds \cite{ref:WMAP}.
Here $S_*$ is the value corresponding to the scale $k_*=0.05$ Mpc$^{-1}$.
We can also compute the scalar spectral index and its running that describe the scale dependence
of the spectrum of the primordial density perturbation $\mathcal{P_R} = (H/\dot{S})^2 ( H/2\pi )^2$ \cite{ref:Perturbatin};
these indices are defined by
\begin{eqnarray}
&&n_s - 1 = \frac{d\ln \mathcal{P_R}}{d\ln k}, \\
&&\alpha_s = \frac{dn_s}{d\ln k}.
\end{eqnarray}
These are approximated in the slow-roll paradigm as
\begin{eqnarray}
&&n_s (S) \sim 1 - 6 \epsilon_S + 2 \eta_{SS}, \\
&&\alpha_s(S) \sim 16 \epsilon_S \eta_{SS} - 24 \epsilon_S^2 - 2\xi^2_{(3)},
\end{eqnarray}
where $\xi_{(3)}$ is an extra slow-roll parameter that includes the trivial third derivative of
the potential.
Substituting $S_*$ into these equations, we have $n_{s* } \sim 0.951$ and
$\alpha_{s*} \sim -2.50 \times 10^{-4}$.

Because $n_s$ is not equal to 1 and $\alpha_{s}$ is almost negligible, our model suggests
a tilted power law spectrum.
The value of $n_{s*}$ is consistent with the recent observations;
the best fit of five-year WMAP data using the power law $\Lambda$CDM model is
$n_s \sim 0.951$ \cite{ref:WMAP}.
Finally, estimating the spectrum $\mathcal{P_R}$ in the slow-roll approximation (SRA),
\begin{equation}
\mathcal{P_R}\sim\frac{1}{12\pi^2}\frac{V^3}{\partial V^2},
\end{equation}
we find $\mathcal{P_R}_* \sim 2.36\times10^{-9}$.
This result matches the measurements as well \cite{ref:WMAP, ref:Hayashi1, ref:Hayashi2}. Incidentally speaking, the energy scale $V\sim10^{-10}$ is also within the constrained range obtained by Liddle and Lyth \cite{ref:Liddle}.

In order to study on the angular power spectrum, we need the tensor perturbation (the gravitational wave) spectrum which is given as follows:
\begin{equation}
\mathcal{P}_{\rm grav} = 8 \left( \frac{H}{2\pi} \right)^2 = \frac{2}{3\pi^2}V.
\end{equation}
In SRA, the spectral index of $\mathcal{P}_{\rm grav}$ is given by the slow-roll parameters
$\epsilon$ and $\eta$ as
\begin{equation}
n_{T} = -2\epsilon.
\end{equation}
Using these parameters $TT$ and $TE$ angular spectrum were well fitted to the WMAP data \cite{ref:Hayashi2}.

However, one of the problems we met is that a tachyonic mode inevitably appear from the mass term of modular field $T$. 
Mass squared of the scalars $S$, $Y$, $T$ in GeV$^2$ are obtained by quadratic terms of expansion of potential $V$ as
\begin{eqnarray}
&&M^2 =
\left( \begin{array}{ccc}
m_S^2 & m_{SY}^2 & m_{ST}^2 \\
m_{SY}^2 & m_Y^2 & m_{YT}^2 \\
m_{ST}^2 & m_{YT}^2 & m_T^2
\end{array} \right) \nonumber \\[5pt]
&&\hphantom{M^2} =
\left( \begin{array}{ccc}
4.80 \times 10^{29} &  1.54 \times 10^{33} & -5.88 \times 10^{26} \\[5pt]
1.54 \times 10^{33} &  4.95 \times 10^{36} & -1.19 \times 10^{29} \\[5pt]
-5.88 \times 10^{26} &  -1.19 \times 10^{29} & 2.86 \times 10^{27}
\end{array} \right).
\end{eqnarray}
Then diagonalized mass squared matrix of $S$, $Y$, $T$ in GeV$^2$ is given
\begin{eqnarray}
&&U^{-1} M^2 U =
\left( \begin{array}{ccc}
m_{S'}^2 & 0 & 0 \\[5pt]
0 & m_{Y'}^2 & 0 \\[5pt]
0 & 0 & m_{T'}^2
\end{array} \right) \nonumber \\[5pt]
&&\hphantom{U^{-1} M^2 U} =
\left( \begin{array}{ccc}
2.96 \times 10^{27} & 0 & 0 \\[5pt]
0 & 4.95 \times 10^{36} & 0 \\[5pt]
0 & 0 & -7.85 \times 10^{25}
\end{array} \right).
\end{eqnarray}

It seems to give arise a tachyonic state. However, if we see the higher order terms of expansion, then the terms spontaneously break a symmetry. In this case the modular invariance of the model is broken after the end of inflation. The physical effect of this symmetry breaking will be discussed in following works.

The appearance of the tachyonic state is, however, rather a general feature of superstring inspired supergravity model \cite{ref:Covi}. The problem is investigated radically to consider all angles of situations of No-Scale supergravity, including whether or not the fact is a defect. However, in order to emphasize that the model is still phenomenologically effective, we choose the case of the modular superfield $T$ is trivial and only is put to $T=1,\ \rm{Im}T=0$ in $K,\ W$ and $V$, since the modular superfield $T$ did not play any roles in our former papers. We should be satisfied in our achievements, at this moment, that our model can well explain the inflation and just after the inflation (preheating stage) at phenomenological level. 

\section{Gravitino mass and Evolution of inflaton after the end of inflation}

Now we will investigate the properties of gravitino and other fields. 
First, gravitino mass is given in this case 
\begin{equation}
m_{3/2} = M_P e^{K/2} |W| = 3.16 \times 10^{12} \,\, {\rm GeV},
\end{equation}
where $\hbar = 6.58211915 \times 10^{-25}$ GeV$\cdot$sec and $M_p = 2.435327 \times 10^{18}$ GeV are used.
Now we will concentrate only on the model to investigate the gravitino production after inflation.
If the chaotic potential
\begin{equation}
V = \frac{1}{2} m^2_\phi \phi^2, \label{chao_pot}
\end{equation}
is considered, the equation of motion of the scalar field is given by
\begin{equation}
\ddot{\phi} + \frac{2}{t} \dot{\phi} + m^2_\phi \phi = 0. \label{Reh_Sca_EOM}
\end{equation}

Then the general solution of this differential equation is obtained as
\begin{equation}
\phi (t) = \frac{c_1}{m_\phi t} \sin ( m_\phi t) + \frac{c_2}{m_\phi t} \cos ( m_\phi t). \label{Reh_Gen_Sol}
\end{equation}
Here, by taking limit $t \to 0$, $c_2 = 0$ follows, and if the amplitude $\bar{\phi}(t)$ is defined 
\begin{equation}
\bar{\phi} (t) \equiv \frac{c_1}{m_\phi t},
\end{equation}
then solution is damping oscillation
\begin{equation}
\phi (t) = \bar{\phi} (t) \sin ( m_\phi t) \label{Reh_Gen_Sol2}.
\end{equation}
In our model, by expanding $V$ around the minimum of $S(t)$, $Y(t)$ and
 fixed $T=1$, and by providing $S(t)$ and $Y(t)$ are real, then we obtained $S(t)$, $Y(t)$ as follows:
\begin{eqnarray}
&&S(t)=S_{{\rm min}}+\sqrt{\frac{8}{3}}\frac{\sin[m_S t]}{m_S t}, \\
&&Y(t)= \frac{1}{\eta^2 (1) e^{1/3} c} e^{-\frac{S(t)}{3b}}.
\end{eqnarray}
These formula are frequently used in the following calculations.

\section{F-term supersymmetry breaking mechanism}

It is proved that the evolutions of $S$ breaks supersymmetry after the end of inflation.

In order to argue on the evolution of $F-$terms, the supersymmetry breaking scale $\alpha$ is defined by Kallosh et al. \cite{ref:Kallosh} as follows:
\begin{eqnarray}
&&\alpha = \left|\dot{\phi} \right|^2+m^iG^{-1}{}^{j}_{i} m_j+V_D \nonumber \\
&&\hphantom{\alpha} = 3{M_p}^{2}(H^2+{m_{3/2}}^{2}),
\end{eqnarray}
where $V_D$ is contribution of $D-$term to the potential, and $m^i$ is
\begin{eqnarray}
m^i \equiv {\cal D}^i m = e^{K/2} \left[ \partial^i W + ( \partial^i K ) W \right], \qquad\qquad
m \equiv e^{K/2} W.
\end{eqnarray}
The $F-$term supersymmetry breaking scale is also given by Nilles et al. \cite{ref:Nilles1,  ref:Nilles2, ref:Nilles3}
\begin{eqnarray}
{f_{\phi_i}}^2\equiv {m_i}^2+\frac{1}{2}\left(\frac{d\phi_i}{dt} \right)^2,
\end{eqnarray}
${f_{\phi_i}}^2$ give a "measure" of the size of the supersymmetry breaking provided by the $F-$term associated with the $i$-th scalar field, which is the same as $\alpha$. The factor $\frac{1}{2}$ in front of $\dot{\phi_i}$ shows $\dot{\phi_i}$ are real. 
In our model these quantities are estimated by calculating  
$m^SG^{-1}{}^{S}_{S} m_{S}$, $m^YG^{-1}{}^{Y}_{Y} m_{Y}$ and time derivatives of these fields, 
finally ${f_{\phi_i}}^2$ are given by
\begin{eqnarray}
f_S^2 = 1.74 \times 10^{25}, \qquad\qquad f_Y^2 = 1.15 \times 10^{21},
\end{eqnarray}
where numerical values are estimated at the stationary points $S_{{\rm min}}$, $Y_{{\rm min}}$ which are also asymptotic values.

By inserting the stationary values of $S$, $Y$ and 
define $\alpha_S={f_S'}^2$, $\alpha_Y={f_Y'}^2$, we obtain the ratios $r_S$, $r_Y$
\begin{eqnarray}
&&r_S=\frac{\alpha_S}{\alpha_S+\alpha_Y}=0.999, \\[5pt]
&&r_Y=\frac{\alpha_Y}{\alpha_S+\alpha_Y}=6.63 \times 10^{-5},
\end{eqnarray}
Thus, supersymmetry is overwhelmingly broken by superfield $S$. Contrary to the fact pointed out by Nilles et al. \cite{ref:Nilles1, ref:Nilles2, ref:Nilles3}, that the interchange of supersymmetry breaking fields occurs,  it does not occur in our model.
The evolution of the ratios is shown at Fig.1.
\begin{figure}[!htbp]
\begin{center}
\includegraphics[width=7cm]{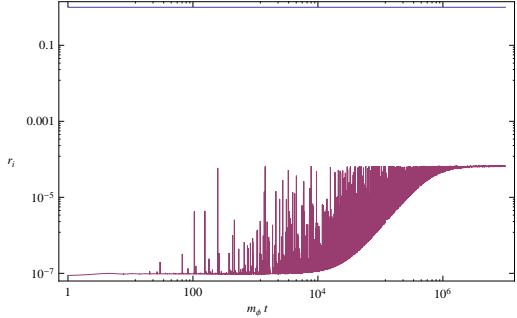}
\caption{Relative contribution of the two scalar fields $S$ and $Y$ to the supersymmetry breaking during their evolution.
That of $r_S$ corresponds to the highest curve, $r_Y$ the lowest.
}
\label{fig:one}
\end{center}
\end{figure}

\section{Goldstino state and Inflaton decay into Gravitino}

Here we will show the value of Fermion masses.
The masses of Goldstino $\tilde{S}$ and supersymmetric partners $\tilde{Y}$ are obtained as follows.
Using $m^{ij} = m^{ji}$ and $\chi_i \chi_j = - \chi_j \chi_i \,\, (i \neq j)$, we obtain: 
\begin{eqnarray}
m^{ij} \chi_i \chi_j 
= m^{\tilde{S}\tilde{S}} \chi_{\tilde{S}} \chi_{\tilde{S}} 
+ m^{\tilde{Y}\tilde{Y}} \chi_{\tilde{Y}} \chi_{\tilde{Y}}, 
\end{eqnarray}
where we have neglected the Hermite conjugate terms. Then, the values are
\begin{eqnarray}
m_{\tilde{S}} = 0 \,\, {\rm GeV}, \qquad\qquad
m_{\tilde{Y}} = 3.68 \times 10^{17} \,\, {\rm GeV}.
\end{eqnarray}
Since $\tilde{S}$ is massless and $S$ breaks supersymmetry, $\tilde{S}$ state is identified with Goldstino, which is absorbed into gravitino by super-Higgs mechanism \cite{ref:SUGRA, ref:moroi}. $\tilde{Y}$ is the supersymmetric partners of $Y$.

Following Endo et al. \cite{ref: Endo, ref:nakamura}, the decay rate 
$\Gamma(S\rightarrow 2\psi_{3/2})$ is calculated in our model:
\begin{equation}
\Gamma_{(S\rightarrow 2\psi_{3/2})}\simeq \frac{1}{288\pi}\frac{|G_S|^2}{G _{S \bar{S}}}\frac{{m_S}^5}{{m_{3/2}}^2{M_p}^2}, \label{decay}
\end{equation}
because the gravitino production comes only from inflaton $S$.
$G_S$ and $G_{S\bar{S}}$ are respectively well defined (by assuming the fields real) at stationary points of three scalars $S$,$Y$ and $T$.
The mass parameter included in $\Gamma_{(S\rightarrow 2\psi_{3/2})}$ is used as:
\begin{eqnarray}
\begin{array}{ccc}
m_S=5.44 \times 10^{13} \,\, {\rm{GeV}}.
\end{array}
\end{eqnarray}
Now the value of the decay rate Eq.(\ref{decay}) is given by
\begin{eqnarray}
&&\Gamma_{(S\rightarrow 2\psi_{3/2})} = 1.56 \times 10^{4} \,\, {\rm{GeV}}.
\end{eqnarray}
The decay time $\tau_{(S\rightarrow 2\psi_{3/2})}$ is \\[-25pt]
\begin{eqnarray}
&&\tau_{(S\rightarrow 2\psi_{3/2})}=\frac{\hbar}{\Gamma_{(S\rightarrow 2\psi_{3/2})}}\simeq 4.23 \times 10^{-29} \,\, {\rm{sec}}.
\end{eqnarray}
This process occurs almost instantly. Thus it is quite possible that primordial gravitino production occurs very rapidly.

\section{Instant preheating temperature in NMSSM}
Now we should choose a model to determine the preheating temperature by inflaton decay into the Minimal Supersymmetric Standard Model (MSSM) or NMSSM particles. We will assume the instant preheating mechanism \cite{ref:fkl}, because this method is mathematically easier to control than that of parametric resonance \cite{ref:kls}.

It is not unique to take in the ordinary particles into No-scale supergravity. Therefore we will assume that the effective superpotential during the oscillations of inflaton is given as follows:
\begin{eqnarray}
W_{NMSSM}= M_SS^2 + M_{R_i}N^c_iN^c_i + \lambda_iSN^c_iN^c_i + \gamma^{ij}_{\nu}N^c_iL_jH_u,
\end{eqnarray}
where $S$ is dilaton (inflaton), $N^c_i$ are the gauge singlet right handed neutrinos, $L_i$ are the lepton doublet superfields and $H_u$ is the Higgs superfield.
In this letter, we will apply the instant preheating mechanism \cite{ref:fkl, ref:fkn} in order to estimate the effects of preheating.
After single oscillation of inflaton, the number density $n_k$ is given by
\begin{eqnarray}
n_k=\exp \left(-\frac{\pi (k^2/a^2+M_{R_i})}{\lambda_i|\dot{S}_0|}\right).
\end{eqnarray}
By integrating to obtain the number density $n_k$ of right handed sneutrinos $\tilde{N}^c_i$:
\begin{eqnarray}
&&n_{\tilde{N}^c_i} = \frac{1}{2\pi^2} \int_0^{\infty} dk k^2n_k \\
&&\hphantom{n_{\tilde{N}^c_i}} = \frac{(\lambda_i\dot{S}_0)^{3/2}}{8\pi^3} \exp \left(-\frac{\pi M^2_{R_i}}{\lambda_i|\dot{S}_0|}\right) \\
&&\hphantom{n_{\tilde{N}^c_i}} \approx \frac{(\lambda_i\dot{S}_0)^{3/2}}{8\pi^3},
\end{eqnarray}
where we concentrated on the case $M^2_{R_i} \lesssim \lambda_i |\dot{S}|$. Thus, the preheating temperature is obtained just after thermal equilibrium:
\begin{eqnarray}
&&T_R = \left(\frac{30}{\pi^2g_*}\cdot m_{\tilde{N}^c_i}\cdot n_{\tilde{N}^c_i}\right)^{1/4} \nonumber \\
&&\hphantom{T_R} = \left( \frac{15 M_{R_i} (\lambda_i\dot{S}_0)^{3/2}}{4 g_* \pi^5} \right)^{1/4} \nonumber \\
&&\hphantom{T_R} = 2.33 \times 10^{12} \,\, {\rm{GeV}},
\end{eqnarray}
%\begin{eqnarray}
%&&T_S=\left(\frac{\pi^2{\rm{g_*}}}{90}\right)^{-\frac{1}{4}}\sqrt{M_p\Gamma_{(S\rightarrow 2\psi_{3/2})}}\simeq 8.70 \times 10^{10} \,\, {\rm{GeV}},
%\end{eqnarray}
%where ${\rm{g_*}}$ is ``effective number of massless degrees of freedom".
where the constants are chosen as $\lambda_i=8.44057\times 10^{-6},\dot{S}_0=8.77073\times 10^{31},M_{R_i}=\frac{1}{2}m_S$ and ${\rm{g_*}}=228.75$ in the case of MSSM. 
By using the formula given in Ref.\cite{ref:takahashi}, the yield variable for gravitino $Y_{3/2}=\frac{ n_{3/2}}{s}$, where $n_{3/2}$ is the number density of gravitino and $s$ is the entropy density of matters, is estimated to 
\begin{eqnarray}
&&Y_{3/2}^{(NT)} = 2\frac{\Gamma_{(S\rightarrow 2\psi_{3/2})}}{\Gamma_\phi}\frac{3T_R}{4m_S} \nonumber \\
&&\hphantom{Y_{3/2}^{(NT)}} = 2\frac{1}{288\pi}\frac{|G_S|^2}{{\rm{G}}_{S\bar{S}}}\frac{m_S ^5}{m_{3/2} ^2 M_P ^2}\left(\frac{10}{\pi^2 {g_*}} \right)^{\frac{1}{2}}\frac{M_P}{T_R ^2}\frac{3T_R}{4m_S} \nonumber \\
&&\hphantom{Y_{3/2}^{(NT)}} = \frac{1}{192\pi}\frac{|G_S|^2}{{\rm{G}}_{S\bar{S}}}\frac{m_S ^4}{m_{3/2} ^2 M_P T_R}\left(\frac{10}{\pi^2 {g_*}} \right)^{\frac{1}{2}} \nonumber \\
&&\hphantom{Y_{3/2}^{(NT)}} = 2.99 \times 10^{-5}.
\end{eqnarray}
The obtained value seems too large at a first glance. However, the produced gravitinos decay rather very rapid, for instance, the decay time of gravitino to a photon and a photino $\psi_{3/2} \rightarrow \gamma + \tilde{\gamma}$ is calculated for $m_{\tilde\gamma}\ll m_{3/2}$ as:  
\begin{eqnarray}
&&\tau_{3/2} (\psi_{3/2} \rightarrow \gamma + \tilde{\gamma}) = 3.9 \times 10^8 \left( \frac{m_{3/2}}{100 \,\, {\rm GeV}} \right)^{-3} \nonumber \\
&&\hphantom{\tau_{3/2} (\psi_{3/2} \rightarrow \gamma + \tilde{\gamma})} = 1.2 \times 10^{-21} \,\, {\rm sec},
\end{eqnarray}
where we have used the formula derived in Ref.\cite{ref:kawasaki-moroi}. 
Because the primordial gravitinos decay very rapidly and the preheating temperature is lower than the gravitino mass,  the effect to the standard Big Bang Nucleosynthesis (BBN) scenario \cite{ref:Kawasaki, ref:kawasaki2}  (see also Refs.\cite{ref:riotto, ref:Buchmuller}) may be negligible in our model.

\section{Conclusion}

We have investigated on the gravitino production just after the end of inflation through the inflaton (dilaton) decay. 
The model we used, cleared the $\eta$-problem and appeared to predict successfully the values of observations at inflation era.
It predicted for examples, the indices $n_{s* } \sim 0.951$ and
$\alpha_{s*} \sim -2.50 \times 10^{-4}$. The value of $n_{s*}$ is consistent with the recent observations; the best fit of five-year WMAP data using the power law $\Lambda$CDM model is
$n_s \sim 0.951$ \cite{ref:WMAP}.
The estimation of the spectrum was as $\mathcal{P_R}_* \sim 2.36\times10^{-9}$, which
result matches the measurements as well \cite{ref:WMAP, ref:Hayashi1, ref:Hayashi2}.  
 
Because the mass of gravitino is calculated as $3.16 \times 10^{12}$GeV, it is rather heavy and may be unstable, therefore, may not be considered as the lightest supersymmetric particle (LSP) or the next lightest (NLSP) and not a dark matter candidate discussed in Refs.\cite{ref: Endo, ref:nakamura, ref:Pradler}. 
However the main topic of supergravity at present stage of the theory is whether the gravitino exist or not in nature despite its mass. 
On the other hand, the supersymmetry breaking is triggered by $F-$term of the inflaton (dilaton) superfield and the interchange of supersymmetry breaking fields does not occur in our model.  Supersymmetry is overwhelmingly broken by superfield $S$ only, contrary to the fact of the interchange of supersymmetry breaking fields pointed out by Nilles et al. \cite{ref:Nilles1, ref:Nilles2, ref:Nilles3}.
From the masses of inflaton $S$ and gravitino mass $m_{3/2}$, the decay rate of the inflaton to two gravitinos is  estimated to $1.56\times 10^{4} {\rm GeV}$. The gravitinos are produced almost instantly just after the end of inflation.
The preheating temperature is also estimated by assuming the instant preheating mechanism and by using a tentative model among NMSSM models \cite{ref:fkl} in order to calculate the entropy density.  Then the yield variable for gravitino takes rather large value, however the decay time appears very rapid and disappear before the BBN stage of the universe.
Therefore we conclude that our present model seems consistent with the present situation of observations for gravitino production and NMSSM matters.

Nonetheless, one of the problems we met is that a tachyonic mode inevitably appear from the mass term of modular field $T$, though we explained the fact by occurrence of spontaneous breakdown of modular invariance. Whether the fact is a defect or not, we need more deeper studies,  however, we emphasize that the model still seems phenomenologically effective.
The tachyonic state is rather a general feature of superstring inspired supergravity model \cite{ref:Covi}. 
We should find satisfaction in our achievements, at this moment, that our model can well explain the inflation and just after the inflation (preheating stage) at phenomenological level. More detailed analysis will appear soon.
Though we have been exclusively restricted our attention to a model of Ref.\cite{ref:Ferrara}, the other models derived from the other type of compactification seems very interesting. Among them KKLT model \cite{ref:Linde, Endo:2005uy, Choi:2005uz, ref:Nilles4} attracts our interest, where the moduli superfield $T$ plays an essential roles. We should take all the circumstances into consideration on essential problems confronted in construction of (No-Scale) Supergravity models.

%%%%%%%%%%%%%%%

\end{document}